\documentclass[useAMS,usenatbib]{mn2e}
\usepackage{graphicx}
\usepackage{amsmath}

\begin{document}

\title{Modeling Increased Metal Production in Galaxy Clusters with Pair-Instability Supernovae}

\author[B. J. Morsony et al.]{Brian J. Morsony,$^{1,2}$\thanks{Email:morsony@astro.wisc.edu} Caitlin Heath$^{3,4}$, Jared C. Workman$^{3}$ \\  
$^1${Department of Astronomy, University of
Wisconsin-Madison, 2535 Sterling Hall, 475 N. Charter Street, Madison WI 53706-1582, USA}\\
$^2${NSF Astronomy and Astrophysics Postdoctoral Fellow}\\
$^3${Dept. of Physical and Environmental Sciences, Colorado Mesa University, Grand Junction, CO, 81501, USA}\\
$^4${APS Department, University of Colorado, Boulder, UCB 391, Boulder, CO 80309, USA}}

\maketitle

\begin{abstract}

Galaxy clusters contain much more metal per star, typically 3 times as much, than is produced in normal galaxies.
We set out to determine what changes are needed to the stellar mass function and supernovae rates to account for this excess metal.
In particular, we vary the Type Ia supernovae rate, IMF slope, upper and lower mass cutoffs, and the merger rate of massive stars.
We then use existing simulation results for metal production from AGB stars, Type Ia SNe and core-collapse SNe to calculate the total amount of each element produced per solar mass of star formation.
For models with very massive stars, we also include metal production from pair-instability supernovae (PISNe).
We find that including PISNe makes it much easier to increase the amount of metal produced per stellar mass.
Therefore a separate population of high-mass stars is not needed to produce the high amounts of metal found in galaxy clusters.
We also find that including at least some PISNe increases the abundance of intermediate-mass elements relative to both oxygen and iron, consistent with observations of ICM abundances.

\end{abstract}


\section{Introduction}
\label{sec:introduction}

The hot intra-cluster medium (ICM) of galaxy clusters is not made of pristine primordial gas, but instead is enriched in iron to $30\%$ solar or more \citep{edge91,yamashita92,mushotzky97}, including super-solar iron abundances in the Centaurus cluster \citep{sanders06,lovisari11}.  
Abundances of other elements are also enhanced.
The ICM in a galaxy cluster typically contains 85\% to 95\% of the total baryons in the cluster, with stars accounting for just 5\% to 15\% \citep{gonzalez07,andreon10}.
The cluster galaxies themselves are slightly enriched relative to field galaxies of the same types \citep[e.g.][]{ellison09}.

Overall, the total amount of metal per star in a galaxy cluster is around 2-6 times more than is found in the Milky way.
Put another way, if all the metal in galaxy clusters were contained in the cluster galaxies, these galaxies would have metallicities 2-6 times solar.

This is very difficult to explain with normal star formation and stellar evolution.
Typically, a galaxy or any collection of stars and gas should saturate at an average metallicity close to solar \citep{tinsley80,pagel97}.
Although some stars will have super-solar metallicity, this is because they formed from gas enriched by previous generations of lower metallicity stars, so the average metallicity is still solar.
For 1 M$_{\sun}$ of stars formed, about $0.02$ M$_{\sun}$ of metal will be produced via nucleosynthesis in AGB stars and supernovae.

Stars not in galaxies, which produce the intra-cluster light, are not sufficient to account for the excess amount of metal if they have a normal IMF, as they make up only about 5\% to 20\% of the stars in clusters \citep{krick07}.

Numerous previous studies \citep[e.g.][]{loewenstein01,portinari04,loewenstein06,loewenstein13} have rejected the notion that the metal in galaxy clusters can come from a standard initial-mass function (IMF).
They concluded that either a separate population of high-mass stars \citep{loewenstein01,loewenstein06} or a top-heavy IMF \citep{portinari04,fabjan08} is necessary to produce enough iron and other metals through additional Type Ia and core-collapse SN.
\citet{bregman10} concluded that the majority of metals in a cluster must form outside of cluster galaxies, in population III stars.

There is observational evidence for an increased rate of Type Ia supernovae in galaxy clusters.  
\citet{mannucci08} and \citet{sharon07} found that Type Ia rates in cluster elliptical galaxies are about 3 times the rate in field ellipticals.
\citep[However, see][]{sand12,maoz12}.
Assuming the increased Type Ia rate reflects a true increase in the number of Type Ia SNe per solar mass of stars formed, it would be insufficient to account for the increased iron amounts in clusters \citep[see sect.~\ref{sec:typeIa},][]{loewenstein06}.
\citet{mannucci08} found the the core-collapse SNe rate in cluster and field galaxies is consistent with being the same, indicating there cannot be a dramatically different IMF in cluster galaxies.

Evidence from low-mass stars also indicates the IMF in galaxy clusters cannot be dramatically different.  
There appears to be an excess number of low-mass stars in large elliptical galaxies \citep[e.g.][]{treu10,auger10,vandokkum10}, possibly indicating a time-variable IMF \citep{weidner13a}.
If there is a large excess population of high-mass stars in clusters, it would have to be above 1 M$_{\sun}$ only, and not just a flattening of the low-mass IMF slope.

Previous studies do not include the impact of vary massive stars ($>140$ M$_{\sun}$) that produce pair-instability supernovae (PISNe).
The IMF is typically truncated at a fixed upper mass of 100 to 150 M$_{\sun}$ \citep{portinari04,loewenstein06,loewenstein13}, approximately the mass of the largest stars in the Milky Way.
Alternately, larger stars could be included but assumed to lose a substantial fraction of their mass and produce core-collapse supernovae.  
In a typical IMF, the number of very massive stars is so small the a few extra core-collapse SNe will not substantially alter the total metal production.
However, if very massive stars are present and they each produce a PISN, they would produce substantially more metal.  
A PISN can, for example, produce a few 10's of solar masses of iron \citep{heger02,galyam09} rather than a few tenths of a solar mass of iron in a core-collapse SN \citep[e.g.][]{heger10}.

\subsection{Evidence star formation is different in clusters}

There are several reasons to think very high-mass stars may be able to form in galaxy clusters.
Cluster galaxies, particularly cD galaxies, have very large star forming regions, up to several million solar masses \citep[see, e.g.][]{canning10}.
They also have very high star formation rates of 10 to 100 M$_{\sun}$ per year in nearby clusters \citep[e.g.][]{mcnamara89,odea10} and up to 740 M$_{\sun}$ per year at high redshift \citep{mcdonald12}.
Very high-mass stars may only form in very large star forming regions.  
This is particularly true if stellar masses in a star forming region are optimally sampling the IMF, rather than randomly sampling it \citep[see][]{weidner13}.
In the Milky Way, the largest star forming regions do contain the highest-mass stars \citep{weidner13}.
In the local group, the largest stars and the largest number of O stars are found in R136, part of the 30 Doradous complex in the LMC, the largest nearby star forming region \citep{massey98}.

The largest star in R136 is $265$ M$_{\sun}$ ($320$ M$_{\sun}$ initial mass), much larger than any star known in the Milky Way \citep{crowther10}, and well above the mass cutoff assumed in a typical IMF.
It is unknown if this star formed as a single star or from the merger of two massive stars \citep{banerjee12}, but in either case is shows that very high-mass stars can form in very massive star forming regions.

cD galaxies in clusters also have a much larger number of globular clusters than the Milky Way.  
Our Galaxy has about 150-200 globular clusters, while M87, with about 10 times the stellar mass, has about 12,000 \citep{tamura06}.
The specific frequency of globular clusters in cD galaxies is about 20 times the frequency in the Milky Way and 3-6 times the frequency in elliptical galaxies \citep{harris91}.
Globular clusters likely began their lives as massive star forming regions, so something about star formation in cD galaxies must allow proportionally more of these regions to exist over the life of the Universe.
This may indicate the typical star formation region size is always larger in cluster galaxies than in the Milky Way.

Aside from the size of the star forming regions, galaxy clusters, at least in their inner regions, are at very high pressure.  
In the inner 100 kpc, the typical ICM pressure in a galaxy cluster is 5 to 50 times the ISM pressure in the Milky Way \citep[e.g][]{arnaud10}.
This presumably means an increased pressure in star forming regions as well, which could lead to more rapid accretion onto stellar cores and therefore a higher upper mass limit.

Very large star-forming regions may also have a slightly flatter IMF slope.
For example, the Arches star cluster appears to have a slope of $-2.1$ rather than the standard $-2.3$ \citep{espinoza09}, and NGC 3603 likely has an even flatter slope, measured as $-1.74^{+0.62}_{-0.47}$ by \citet{harayama08}.
Although this is only slightly flatter, the slope has the largest effect as the highest masses, so even a small change can significantly increase the number of very high mass stars.

Even if the {\it initial} mass function in cluster galaxies has the same mass cutoff as the Milky Way, very high mass stars may still form.
Galaxies in clusters, and very massive star forming regions, have very dense stellar environments.
This will increase the rate of stellar interactions compared to field galaxies, and a hardening of binary systems.
\citet{sana13} found that 50\% of the O stars in 30 Doradus will exchange mass with a companion during their lifetime.
This could lead to mergers of high mass stars to produce very high mass stars \citep{banerjee12}.
Type Ia supernovae are believed to be produced from binary interactions, either accretion onto a white dwarf from a companion or a merger of two white dwarfs, so an increase in stellar interactions may also explain the higher Type Ia SNe rate seen in cluster galaxies \citep{mannucci08}.

Pressures and interaction rates will be highest in the central regions of galaxy clusters, and particularly in cD galaxies.  
However, this accounts for only a small fraction of the total galaxy cluster mass (about 10\% of stars are in the cD galaxy).
If extra metal is produced only in this region, the excess relative to ordinary star formation would have to be very large, and the metal would need to be transported to the outer region of the cluster.
This could possibly be accomplished through AGN driven outflows or sloshing of cluster gas.
In galaxy clusters without a cD galaxy, or with lower density non-cool cores, it may be less likely to have a significantly modified IMF, and therefore be more difficult to produce excess metals.

If enhanced metal production compared to the Milky Way is a general feature of elliptical galaxies, and not just galaxies in the inner regions of clusters, we might expect evidence from observations of metals in ellipticals and in the IGM, particularly if PISN are responsible.
Stellar abundances in the centers of ellipticals, determined through optical observations of integrated starlight, indicate that the total amount of metal is enhanced relative to solar by a factor about 2, and that $\alpha$ elements (typically Mg) are enhanced relative to iron by about 1.5 \citep[see, e.g.,][]{trager00, trager00b, tang09}.
These enhancements also correlate with galaxy mass.
The IGM, though far below solar metallicity, has an apparent enhancement of Si relative to carbon ([Si/C]$=0.77 \pm 0.05$ rather than $\approx 0.5$), possibly indicating a significant contribution from PISNe \citep{aguirre04}.

We aim to explore what combination of stellar mass function and supernovae rates, including PISNe, are needed to produce the high amount of metal per star seen in galaxy clusters.
In section~\ref{sec:base_models} we explain our methods for calculating total metal production and compare our results using standard IMF base models to observed solar metallicity values.
In section~\ref{sec:results} we explore how various modifications to our base models increase metal yield.
In section~\ref{sec:compare_obs} we compare the results of our models to observed abundances of various elements in galaxy clusters.
In section~\ref{sec:kobayashi} we examine how our results change if we use an alternative set of core-collapse supernova models.
Finally in section~\ref{sec:conclusions} we summarize our conclusions.

\section{Modeling Metal Production}
\label{sec:base_models}

To determine the abundances of different elements produced, we start with 1 M$_{\sun}$ of star formation for a given initial-mass function (IMF), and then determine the number of AGB stars and different types of supernova in a given mass range that will be produced.
Results from simulations of metal production are then used to find the total amount of each element produced.  
We can then compare the results of various models to two base models representative of normal, Milky-Way-like star formation.

\begin{figure*}
\includegraphics[scale=.9]{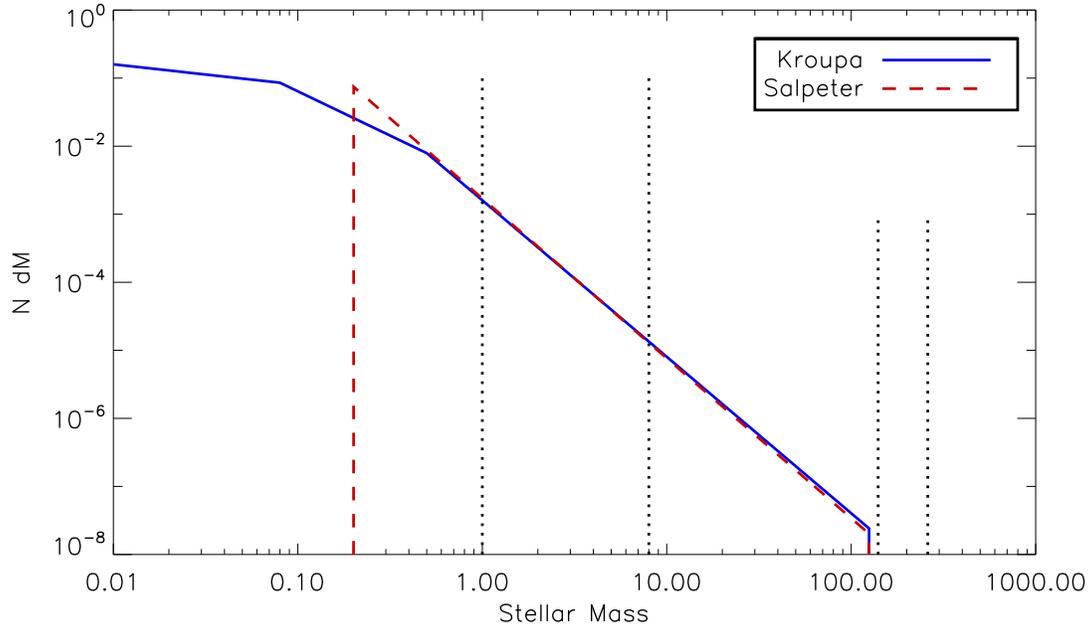}
\caption{
The number of stars vs. stellar mass for the Kroupa IMF (solid blue line) and Salpeter IMF (dash red line), used as our base models. 
AGB stars and Type Ia supernova are assumed to occur only between 1 and 8 M$_{\sun}$ (first and second dotted black lines), core-collapse supernovae between 8 and 140 M$_{\sun}$ (second and third lines), and PISNe between 140 and 260 M$_{\sun}$ (third and fourth lines).
Stars above 260 M$_{\sun}$ are assumed to collapse directly to a black hole with no contribution to metal production.
}
\label{fig:solar_imf}
\end{figure*}

For our bases cases, we use standard Salpeter and Kroupa IMFs (Salpeter 1955, Kroupa 2001).  
The Salpeter IMF is a powerlaw with a slope of $-2.35$, a low-mass cutoff at $0.2$~M$_{\sun}$ and a high-mass cutoff at $125$~M$_{\sun}$.
The Kroupa IMF consists of 3 connected powerlaws with slopes of $-0.3$, $-1.3$ and $-2.3$, with breaks at $0.08$~M$_{\sun}$ and $0.5$~M$_{\sun}$.
There is a low-mass cutoff at $0.01$~M$_{\sun}$ and a high-mass cutoff at $125$~M$_{\sun}$ for our base model.
Figure~\ref{fig:solar_imf} shows the Salpeter and Kroupa IMFs we use for our base models.

Beginning with these IMFs, we assume that all stars between $1$~M$_{\sun}$ and $8$~M$_{\sun}$ become AGB stars. 
We also assume that $1\%$ of stars in this mass range produce a type Ia supernova.  
This produces a rate of 1 type Ia supernova for every 7 type II supernova, over the life of the Universe.
This rate is chosen to produce the approximately correct amount of iron in our base models (see below).

All stars between $8$~M$_{\sun}$ and $140$~M$_{\sun}$ are assumed to produce core-collapse supernovae.
For our base models, there are no stars above $125$~M$_{\sun}$, so all metal production comes from AGB winds, type Ia and core-collapse supernovae.
However, for other models that include higher-mass stars, each star between $140$~M$_{\sun}$ and $260$~M$_{\sun}$ produces a pair-instability supernova (PISN).
Above $260$~M$_{\sun}$, stars are assumed to collapse directly into a black hole, and not contribute at all to metal production.

Metal production for AGB stars is taken from \citet{karakas10}. 
At a given mass, we add equal contributions from models with metallicity fractions of 0.0001, 0.004, 0.008 and 0.02 and normal (Reimer's) mass loss rates.  
We include AGB models from 1.0 to 6.0 M$_{\sun}$  from \citet{karakas10}.
All stars between 1.0 and 8.0 M$_{\sun}$ are assumed to go through an AGB phase, with the amount of each elements produce smoothly interpolated between model masses from 1.0 to 6.0 M$_{\sun}$ and assumed to equal the 6.0 M$_{\sun}$ model between 6.0 and 8.0 M$_{\sun}$.

Metal production for core-collapse supernovae comes from \citet{heger10}.
At a given mass, we add equal contributions from models with supernova energies of 0.3, 0.6, 0.9, 1.2, 1.5, 2.4, and $3.0 \times 10^{51}$~erg.
For all models we assume no mixing and a piston at $S/N_A k_b = 4$.
Initial masses of the models range from 10.0 to 100.0 M$_{\sun}$, with the amount of each elements produce smoothly interpolated between these model masses and assumed to equal the 10.0 M$_{\sun}$ model between 8.0 and 10.0 M$_{\sun}$ or the 100.0 M$_{\sun}$ model between 100.0 and 140.0 M$_{\sun}$.

The core-collapse progenitors from \citet{heger10} are all initially metal free.  
Most stars in galaxy cluster will not be forming in metal-free environments, but these are the best available supernova models that cover a wide range of initial mass and supernova energy.
They also lead an enrichment close to solar metallicity for a standard IMF (see sect.~\ref{sec:base_models_numbers} below).
The best alternative models for core-collapse supernova yields are from \citet{kobayashi2006}, which cover initial metallicity fractions of 0.0, 0.001, 0.004 and 0.02, but have a limited mass range and energy range.
However, even \citet{kobayashi2006} models with similar initial mass, metallicity and supernova energy produce significantly difference yields than in \citet{heger10}.
We consider briefly how using the \citet{kobayashi2006} models would change our results in sect.~\ref{sec:kobayashi}.

Metal production for type Ia SNe is taken from \citet{iwamoto99}.
We use yields from model W7 for all type Ia supernovae.
Metal production for PISNe comes from \citet{heger02}.
Initial masses of the models range form 140 to 260 M$_{\sun}$.

\subsection{Base Models}
\label{sec:base_models_numbers}

Using these choices of AGB and supernova models and type Ia rate, we begin with a standard Salpeter or Kroupa IMF and assume all stars larger than 1.0 M$_{\sun}$ have gone through an AGB phase or become a supernova, and the the correct fraction have produced at type Ia supernova.
We then divide by the initial mass of stars formed to find the final (saturated) metallicity.
For our models, we do not consider the history of metal enrichment, aside from using a fixed range of metallicities for AGB stars.
We assuming no mixing with additional primordial gas, and that gas recycling does not change the metal yield per solar mass of stars formed.

\begin{figure*}
\includegraphics[scale=.9]{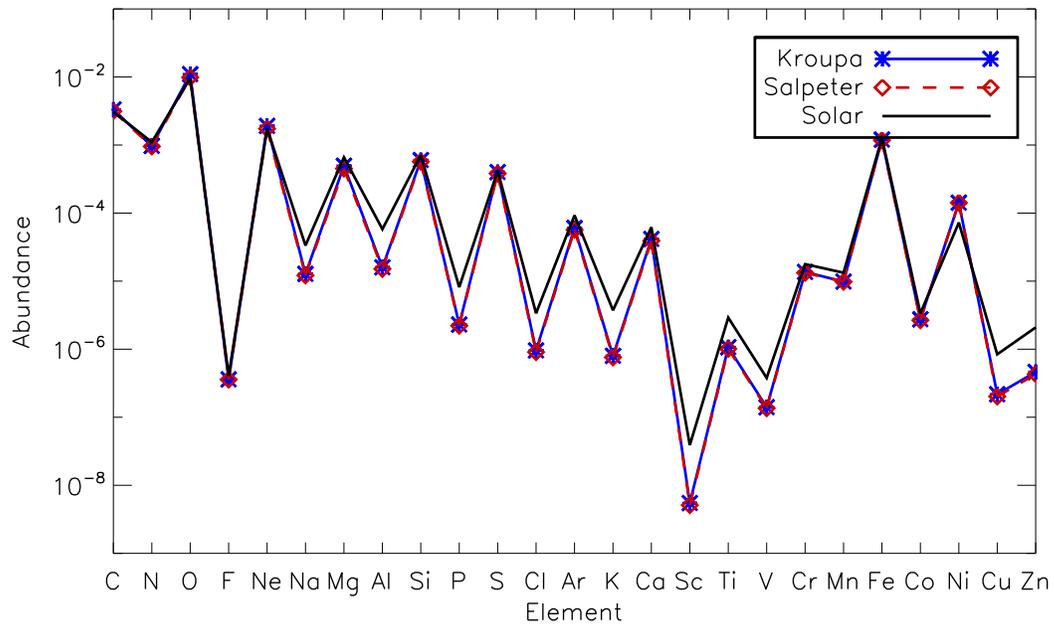}
\caption{
Fractional mass abundance of each element from C to Zn for base models using a Kroupa IMF (solid blue line) and Salpeter IMF (dashed red line).
Also plotted are solar abundances (solid black line) from \citet{anders89}.
}
\label{fig:base_models}
\end{figure*}

\begin{figure*}
\includegraphics[scale=.9]{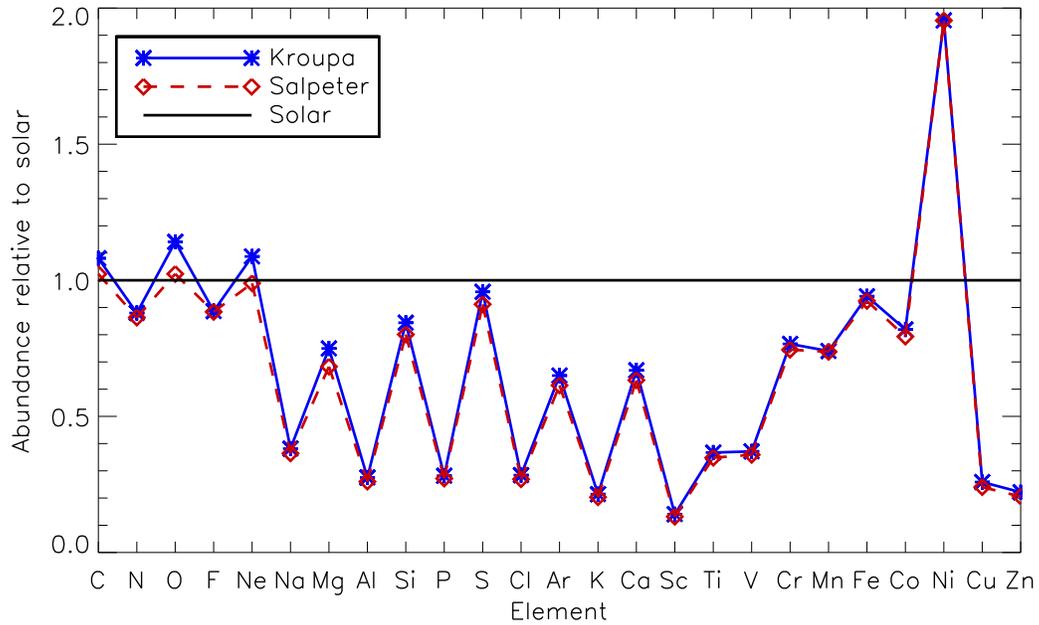}
\caption{
Mass abundance of each elements from C to Zn relative to solar values for base models using a Kroupa IMF (solid blue line) and Salpeter IMF (dashed red line).
Solar abundances (solid black line) are all equal to 1 on this plot.
}
\label{fig:base_vs_solar}
\end{figure*}

We are able to produce metallicities close to solar values.
The amount of each element produced is shown in Fig.~\ref{fig:base_models}, along with solar values from \citet{anders89}.
Figure~\ref{fig:base_vs_solar} shows the amounts of each element produced by the base models for the two IMFs compared to solar values.  
The amount of the most significant elements produced is generally within 30\% of the solar value.
Reasonable values are obtained for the light elements, particularly oxygen, and for iron.
For the even-numbered intermediate elements (Ne, Mg, S, Si, etc.) the amounts produced are a bit low, but still reasonable.
Abundances of odd-numbered intermediate elements are low, only about 30\% of solar values.

The important point, however, is that the standard IMFs produce abundances reasonably close to solar.
For the results in section~\ref{sec:results} we will compare abundances produced by various models to these standard cases, rather than directly to solar abundances, in order to more clearly show how the changes affect metal production.
Keep in mind, however, that the change in abundances for some elements will be overstated, particularly for odd-numbered intermediate elements.

\begin{figure*}
\includegraphics[scale=.9]{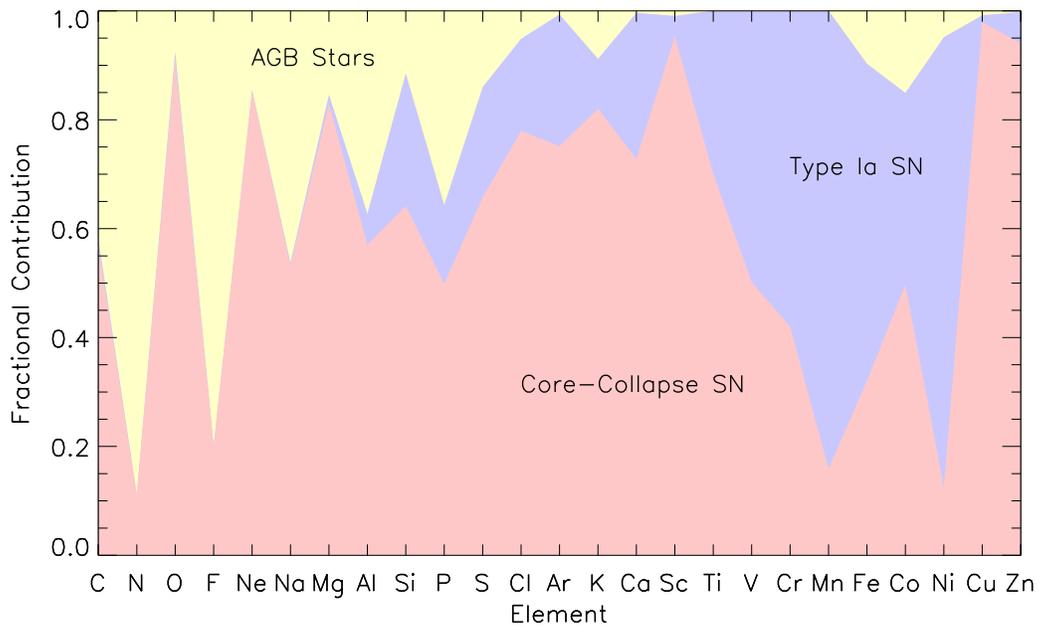}
\caption{
Fraction of each element in the base model with a Kroupa IMF produced by core-collapse SNe (pink), Type Ia SNe (blue) and AGB stars (yellow).
}
\label{fig:base_kroupa_by_type}
\end{figure*}

Figure~\ref{fig:base_kroupa_by_type} shows the contributions of each means of metal production to each elements.  
AGB stars mainly produce the light elements C, N and F, type Ia supernovae mainly produce iron peak elements, and type II supernovae produce most of the oxygen along with the majority of the intermediate elements and about $1/3$rd of the iron.

\section{Results}
\label{sec:results}

To increase the amount of metal produced for a given amount of star formation, we now make various changes to the IMFs and supernova rates.
This includes changing the Type Ia SN rate, the slope of the IMF, removing low-mass stars, extending the high-mass cutoff of the IMF to include pair-instability supernovae, and merging massive stars to produce PISNe.
We also consider various combinations of these changes.
Each model is the compared to the base models to determine the changes in amounts of different elements.

\subsection{Enhanced Type Ia Supernova Rate \label{sec:typeIa} }

One possible way to increase the amount of metal in galaxy clusters is to increase the number of type Ia supernovae.  
Galaxies in clusters could have a higher Ia rate because they generally have older stellar populations, allowing more time for stars to evolve, or because there are more close binaries in clusters.
Galaxies in clusters have, on average, older stellar populations than, for example, the Milky Way.
The total number of Ia supernovae that have occurred should therefore be larger, due to having more time for stars to evolve off the main sequence or for binary mergers to occur.
However, the time-delay distribution of type Ia supernova is skewed towards short delays, so this should not be a large effect.

Type Ia supernovae are believed to be produced by either the merger of two white dwarfs or accretion onto a white dwarf from a companion star.  
In either case, increasing the number of close binaries in the right mass range would increase the type Ia supernova rate.
The high density of stars in cluster galaxies could increase the number of close binaries by hardening through repeated stellar encounters.

Observations of galaxy clusters from \citet{mannucci08} and \citet{sharon07} found a type Ia rate in cluster elliptical galaxies about 3 times the rate in field ellipticals.
However, this result has been challenged by \citet{sand12} and may not be reliable (see \citet{maoz12} for review).

\begin{figure*}
\includegraphics[scale=.9]{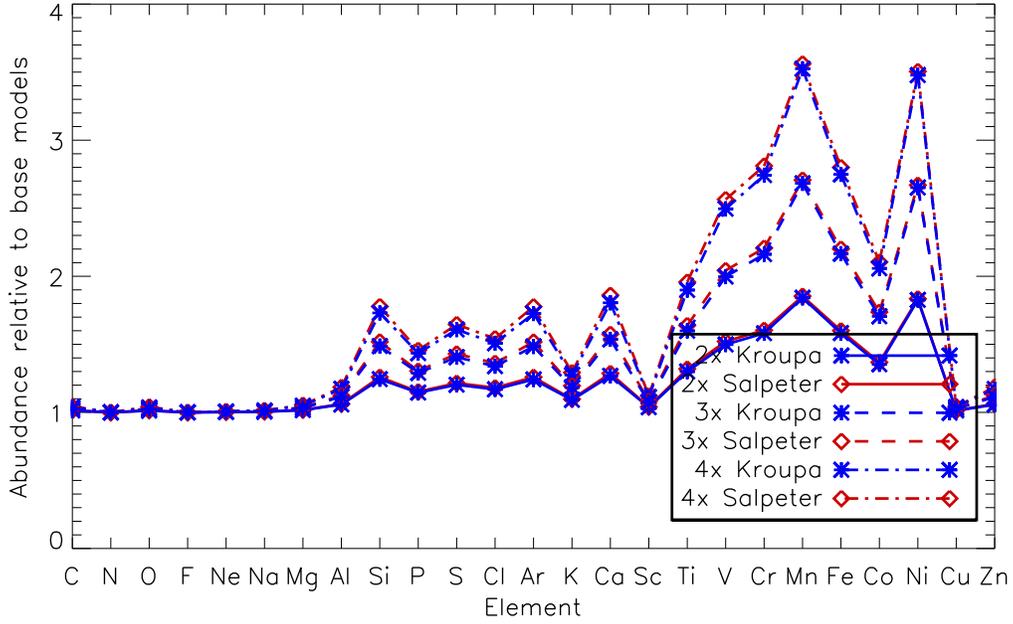}
\caption{
Mass abundance relative to base models for increases in Type Ia SNe rate of 2, 3 and 4 times.  
Models with Kroupa IMFs are blue lines and with Salpeter IMFs are red lines.
There is a large increase in iron and other heavy elements, a moderate increase in intermediate elements and almost no increase in light elements.
}
\label{fig:typeIa_vs_base}
\end{figure*}

In fig.~\ref{fig:typeIa_vs_base} we plot the increase in abundances of each element for a type Ia SNe rate of 2\%, 3\%, and 4\% of stars between 1.0 and 8.0 M$_{\sun}$, relative to the base cases of 1\%.  
Increasing the type Ia rate leads to a clear increase in the iron abundance (58\%, 117\% and 175\%, respectively), but only a slight increase in intermediate elements and almost no increase in light elements.
Although an increased type Ia rate in galaxy clusters could greatly increase the amount of iron present, it cannot explain a significant increase in lighter elements.
It would also lead to a depletion of light and intermediate elements relative to iron.

\subsection{Modifying the IMF slope \label{sec:IMF} }

Additional metal can also be produced by using an IMF with a flatter slope, leading to more high-mass stars and more supernovae and AGB stars for a given amount of star formation.
It is possible stars in galaxy clusters form with a flattened IMF.
Something about either the size of the star-forming regions in galaxy clusters or environmental conditions within the cluster could lead to a flattened IMF.
For example, if the stars typically form in massive star-forming regions, they could form with a flatter IMF.
However, an extremely flat slope to the IMF would lead to an observable deficiency of low-mass stars relative to solar mass and larger stars.
This is inconsistent with cluster observations \citep[e.g.][]{treu10,auger10,vandokkum10} which show an over-abundance of low-mass stars.

\begin{figure*}
\includegraphics[scale=.9]{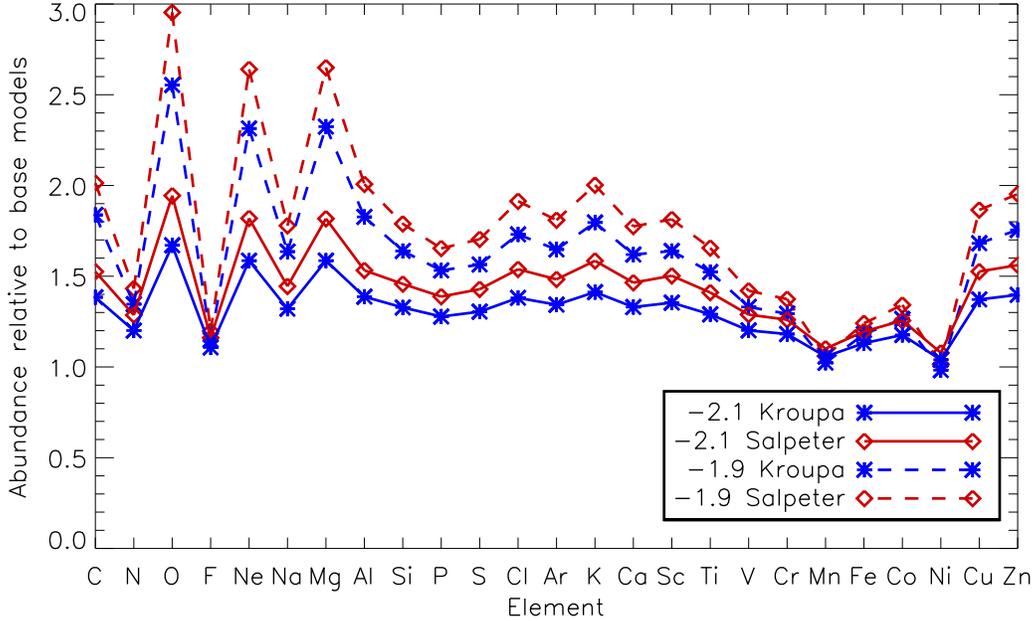}
\caption{
Mass abundance relative to base models for flattened IMF slopes of $-2.1$ and $-1.9$.  
Models with Kroupa IMFs are blue lines and with Salpeter IMFs are red lines.
There is a large increase oxygen abundance, a moderate increase in intermediate elements and only a small increase in iron production.
There is a larger increase for Salpeter models because the slope in the base case is steeper at $-2.35$ vs. $-2.3$ for the Kroupa base model.
}
\label{fig:flat_imf_vs_base}
\end{figure*}

In fig.~\ref{fig:flat_imf_vs_base} we plot the increase in abundances of each element for IMF slopes of $-2.1$ and $-1.9$, relative to the standard IMF cases with slopes of $-2.3$ (Kroupa) and $-2.35$ (Salpeter).  
Flattening the IMF can lead to a significant increase in the amount of metals produced, particularly oxygen, which increases by 67\% and 155\%, respectively.  
There is also a smaller increase in the production of intermediate elements, and a very small ($\approx 15\%$) increase in the amount of iron.
Although the total mass of metal produced can increase by up to 120\%, mainly due to the production of oxygen, the abundance pattern changes significantly from the standard IMF cases.  
The oxygen to iron ratio increases by about 50\% and 120\% relative to the base cases.
Just by having a flatter IMF slope in galaxy clusters it would be difficult to produce enough iron to match cluster observations.  
There would also be a increase in the abundances of light and intermediate elements relative to iron, and in the ratio of oxygen to intermediate elements.

\subsection{Removal of low-mass stars \label{sec:lowmass} }

\begin{figure*}
\includegraphics[scale=.9]{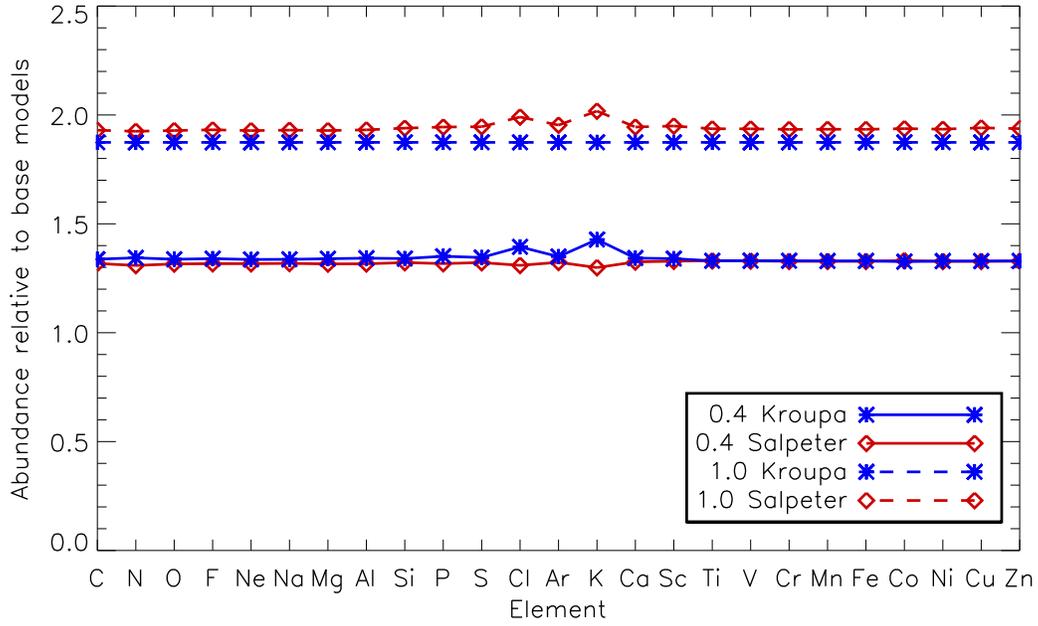}
\caption{
Mass abundance relative to base models for low-mass cutoffs at 0.4 and 1.0 M$_{\sun}$.  
Models with Kroupa IMFs are blue lines and with Salpeter IMFs are red lines.
The relative abundances stay constant for all models, with only an increase in the total metal production.
}
\label{fig:remove_low_mass_vs_base}
\end{figure*}

Decreasing the number of low-mass stars in the IMF will increase the total amount of metals produced without altering their abundances relative to each other, so long as only stars smaller than 1 M$_{\sun}$ are removed.
Figure~\ref{fig:remove_low_mass_vs_base} plots the increase in abundances of each elements for a low-mass cutoff of 0.4 M$_{\sun}$, rather than 0.2 M$_{\sun}$ for the Salpeter IMF and $0.01$ M$_{\sun}$ for the Kroupa IMF.
This results in the amount of metal produced increasing by about 33\%, with no change in relative abundances.  
Even removing all stars smaller than 1 M$_{\sun}$ only doubles the amount of metal produced per stellar mass formed.
However, such a reduction in the number of low-mass stars is not found in galaxy clusters.
Numerous studies \citep[e.g.][]{treu10,auger10,vandokkum10} have found there are more low-mass stars in large elliptical galaxies than expected from a standard IMF, making it very unlikely that metal production in clusters is boosted in this fashion.

\subsection{Pair-Instability Supernovae: Change in high-mass cutoff \label{sec:PISN} }

For both of our standard IMFs, there is a high-mass cutoff in the mass distribution at $125$ M$_{\sun}$, roughly the mass of the largest stars in the Milky Way.
However, this cutoff is not necessarily fundamental.
In the Milky Way, the largest stars are found in the most massive star forming regions \citep{weidner13}.
Larger stars, up to 320 M$_{\sun}$ \citep{crowther10} are found in the Large Magellanic Cloud, which also has the most massive star-forming clouds in the local group.
Galaxies in clusters, particularly cD galaxies, have high star formation rates ($> 10$ M$_{\sun}/$year) and many very large star forming regions ($> 10^6$ M$_{\sun}$), so the high-mass cutoff might be higher in galaxy clusters.

We therefore extend the powerlaw slope of our IMFs up to $400$ M$_{\sun}$.
Although there are still very few stars above $125$ M$_{\sun}$, they make a significant contribution to the amount of metal produced if the create pair-instability supernovae.
The exact minimum mass for a PISN to take place is uncertain, and may be as low as $65$ M$_{\sun}$ \citep{chatzopoulos12}, but we adopt a minimum initial mass of $140$ M$_{\sun}$ for PISN production, taken from \citet{heger02}.
We also adopt a maximum PISN initial mass of $260$ M$_{\sun}$, and assume anything larger than this does not produce a supernova.  
The actual maximum cutoff in the IMF we choose therefore make very little difference, so long at it is greater than $260$ M$_{\sun}$.

\begin{figure*}
\includegraphics[scale=.9]{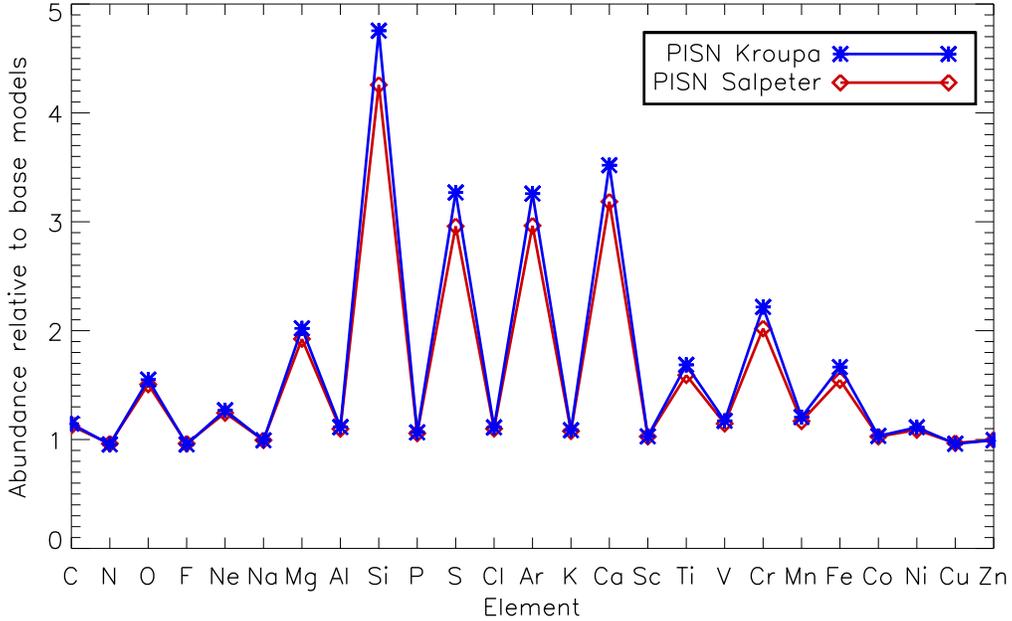}
\caption{
Mass abundance relative to base models for models with upper-mass cutoff of the IMF extended to 400 M$_{\sun}$.  
Star between 140 and 260 M$_{\sun}$ are assumed to produce PISNe.
Model with Kroupa IMFs is the blue line and with Salpeter IMFs is the red line.
There is a very large increase in the production of intermediate elements, along with significant increases in oxygen and iron production (55\% and 67\%).
}
\label{fig:pisn_vs_base}
\end{figure*}

In fig.~\ref{fig:pisn_vs_base} we plot the increase in abundances when PISN are included, compared to the standard IMF cutoff.
Just including the contribution from PISN with a normal IMF slope greatly increases the metal yield.
The amount of O increases about 55\% and the Fe increases about 67\%.
The increase in intermediate mass elements is even more dramatic.
The amount of Si, for example, increases by 475\% and S more than triples.
The total mass of metal produced, dominated by O, increases 58\%.
This alone is not enough to account for the entire amount of excess metal in galaxy clusters, but it is essentially produced for free, without any changes to low-mass star formation or supernova rates.

The presence of PISN will also leave a distinct abundance signature.
The ratio of O/Fe is decreased about 7\% compared to solar values, while the ratio of intermediate elements (Si, S, Ar, etc.) are enriched relative to Fe by a factor of about 2.

\subsection{Massive star mergers \label{sec:merger} }

Very massive stars ($> 125 $ M$_{\sun}$) can also be produced even without increasing the maximum mass cutoff of the IMF if two massive stars collide.
Even if the IMF in galaxy clusters has a standard cutoff of $125$ M$_{\sun}$, the very massive star-forming regions and high stellar densities may lead to more mergers of massive stars.
This is also a plausible way for forming the most massive star in the LMC \citep{banerjee12}.

To test how this effects metal yield, we assume that a fixed fraction of stars above 65 M$_{\sun}$ are in equal-mass binaries and merge, and that the resulting star behaves the same as a star with an initial mass equal to the total of the two merging stars.
Merger products with a mass above $140$ M$_{\sun}$ again produce PISN.
The minimum mass of 65 M$_{\sun}$ is chosen for convenience.
Lowering this mass will change the number and mass of stars that produce core-collapse SNe, but has very little effect on the overall metal production.

\begin{figure*}
\includegraphics[scale=.9]{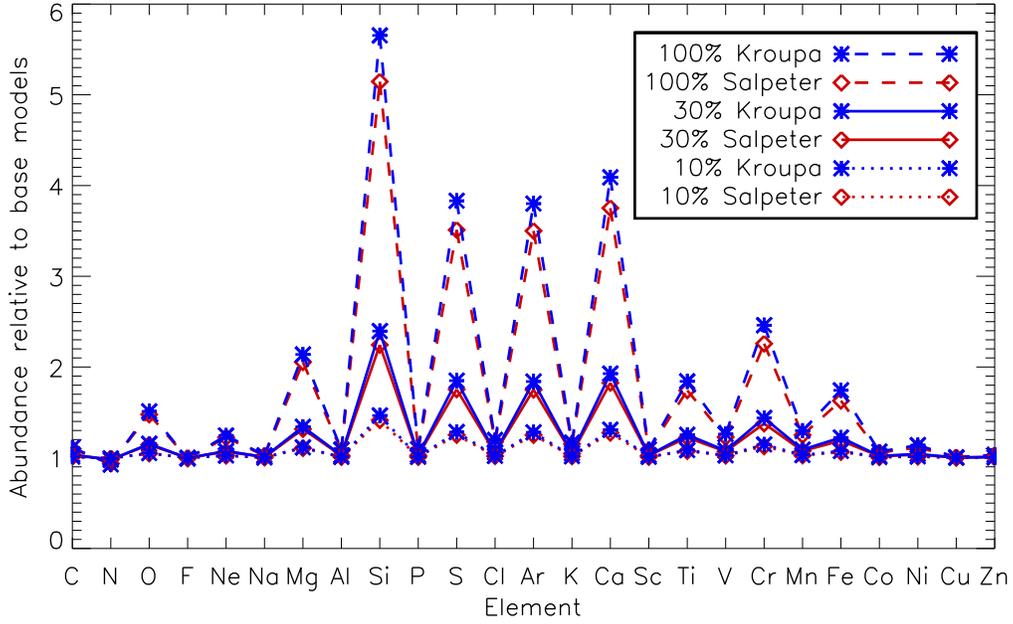}
\caption{
Mass abundance relative to base models for models with mergers of 10\%, 30\% and 100\% of stars above 65 M$_{\sun}$.  
Models with Kroupa IMFs are blue lines and with Salpeter IMFs are red lines.
The 100\% merger case produces similar results to the PISN models (figure~\ref{fig:pisn_vs_base}).
The 10\% and 30\% models produced similar abundance patterns but with more moderate increases in metal production.
}
\label{fig:merger_vs_base}
\end{figure*}

Figure~\ref{fig:merger_vs_base} plots the increase in abundance for merger rates of 10\%, 30\% and 100\%, relative to the standard IMF with no mergers.
The abundance enhancements follow a similar pattern to changing the high-mass cutoff (sect.~\ref{sec:PISN} above) with strong increases in intermediate elements relative to O and Fe, and a slight decrease in the O/Fe ratio.  The excess metal produced is roughly proportional to the merger rate, with the 10\% merger model producing increases about 10\% as much as in the high-mass-cutoff models.
The 100\% merger rate models produce abundance increases about 5\% larger than the high-mass-cutoff model, because the high-mass slope of the IMFs is steeper than 2, so $2\times N(M) > N(2\times M)$.

\subsection{Combined Models \label{sec:combined} }

Making any single change to the IMF or the supernova or merger rates cannot account for the factor of 2-6 increase in metal in galaxy clusters, except in the most extreme cases (very flat IMF at high mass or removal of large number of low-mass stars).
However, combining several changes can lead to greatly increased metal production.

\begin{figure*}
\includegraphics[scale=.9]{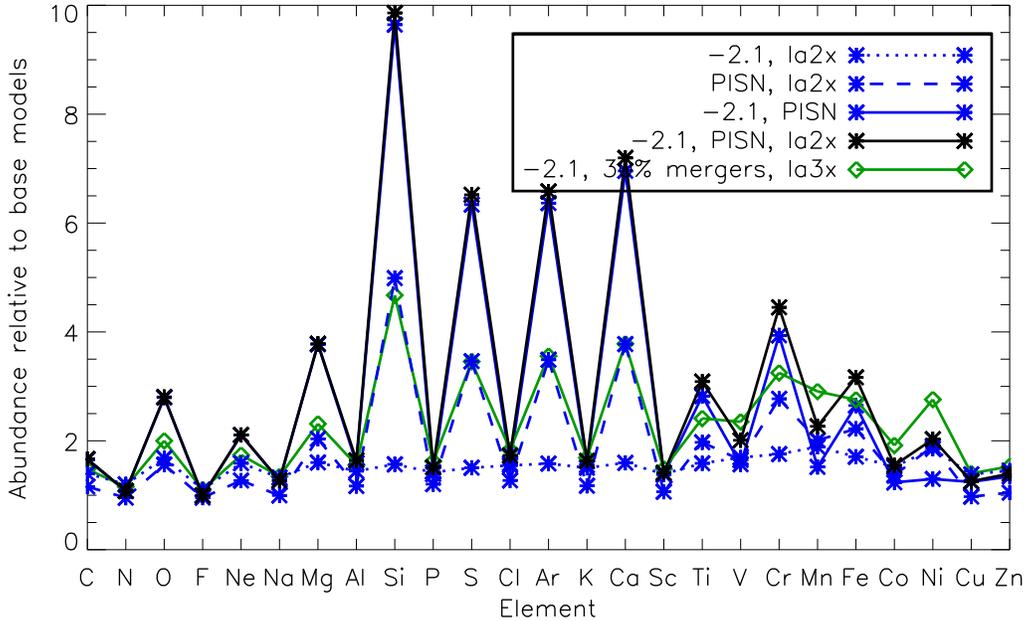}
\caption{
Mass abundance relative to base models for 5 combined models.  
Only models with a Kroupa-type IMF are shown.
All models that include PISNe show a clear signature in the enhancement of intermediate elements.
The largest increases in metal production are those with a flattened IMF and PISN (extended upper mass cutoff).
A compromise model (solid green line) with a flattened IMF, 30\% massive star merger rate, and 3x increased Type Ia SNe rate, produced similar increases in metal production to the PISN, 2x Type Ia rate model, but with more oxygen production.
}
\label{fig:combined_vs_base}
\end{figure*}

In this section, we consider 5 combined models: 1) IMF slope of $-2.1$ and 2x Type Ia rate, 2) 2x Type Ia rate and PISN (high-mass cutoff of $400$ M$_{\sun}$), 3) IMF slope of $-2.1$ and PISN, 4) a combination of all 3 (slope of $-2.1$, 2x Ia rate and PISN), and 5) a slope of $-2.1$, merger of 30\% of stars above 65 M$_{\sun}$ and 3x the Type Ia rate.
Figure~\ref{fig:combined_vs_base} plots the abundance increases for these 5 models relative to the bases cases.

For model 1, the sharp increase in the O/Fe ratio seen in the model with just a flattened IMF slope of $-2.1$ (section~\ref{sec:IMF}) is eliminated, with the ratio now 0.91 compared to the base model.
The abundances of intermediate mass elements relative to O and Fe are only slightly depleted compared to the base case.
However, the total amount of metal produced has still only increased by about 60\%

Model 2, combining PISN and a 2x type Ia SN rate, partially eliminates the depletion of oxygen relative to iron found in the model with just a 2x type Ia rate (section~\ref{sec:typeIa}); it is now only 0.70 rather than 0.64.
The abundances of intermediate elements are significantly enhanced relative to oxygen, similar to the PISN case (section~\ref{sec:PISN}), and relative to iron (up to 224\%).
The total increase in metal production, however, is still only about 64\%.

In model 3, the flattened IMF slope leads to a large increase in the number of PISN produced.
This leads to a huge increase in the metal yield, about 180\% for O, 165\% for Fe and over 9 times as much for some intermediate elements.
The relative abundances follow a similar pattern to the PISN only models (section~\ref{sec:PISN}), with an increase in the O/Fe ratio and a large relative increase in the abundance of intermediate elements.

Model 4 is similar to model 3, but with added iron from the increased type Ia SN rate.
Here the ratio of O/Fe is close the base case (0.89).
There is still a significant enhancement in intermediate elements, with up to a 3 times as much relative to Fe.
The total metal produced is also very high, with oxygen and iron yield of about 3 times the base case.

Model 5 (green line in figure~\ref{fig:combined_vs_base}) produces a more modest increases in abundances that model 4, with, for example, about a 70\% increase in Si/Fe rather than a 210\% increase.
Oxygen is more depleted relative to Fe than in model 4, due to the higher type Ia rate and decreased number of PISN, with an O/Fe abundance of 0.72 relative to the base case.
The total amount of iron produced, however, it still quite high, 2.75 times the base case rather than 3.16 times for model 4.

Models 3, 4 and 5, which all include PISN and a slightly flattened IMF slope at high mass, all produce a total amount of metal per stellar mass formed similar to that seen in galaxy clusters.
Models 3 and 4 also both have relative abundance patterns similar to what is seen in PISN, with very large enhancements in intermediate elements relative to both iron and oxygen (and other low-mass elements).
Model 5 has a lower enhancement of intermediate elements relative to oxygen and iron, but the presence of PISN is still clear in the abundance pattern.

\section{Comparison with Observations}
\label{sec:compare_obs}

We compare the results of our models to observations of observed metal abundances in galaxy clusters.
In particular, we compare to the observations of \citet{lovisari11} of ICM abundances at a range of radii in 5 clusters: Centaurus, A496, Sersic 159-03, Hydra A, and A2029.  
Observations extend to a maximum radius of 180, 350, 390, 500, and 520 kpc, respectively.
They found that generally oxygen is depleted relative to iron in these clusters and that intermediate elements (Si, S, and Ar) are near solar values \citep[from][]{anders89} relative to iron (slightly enhanced in Si, slightly depleted in Ar), but all are significantly enhanced relative to oxygen.

\begin{figure*}
\includegraphics[scale=.9]{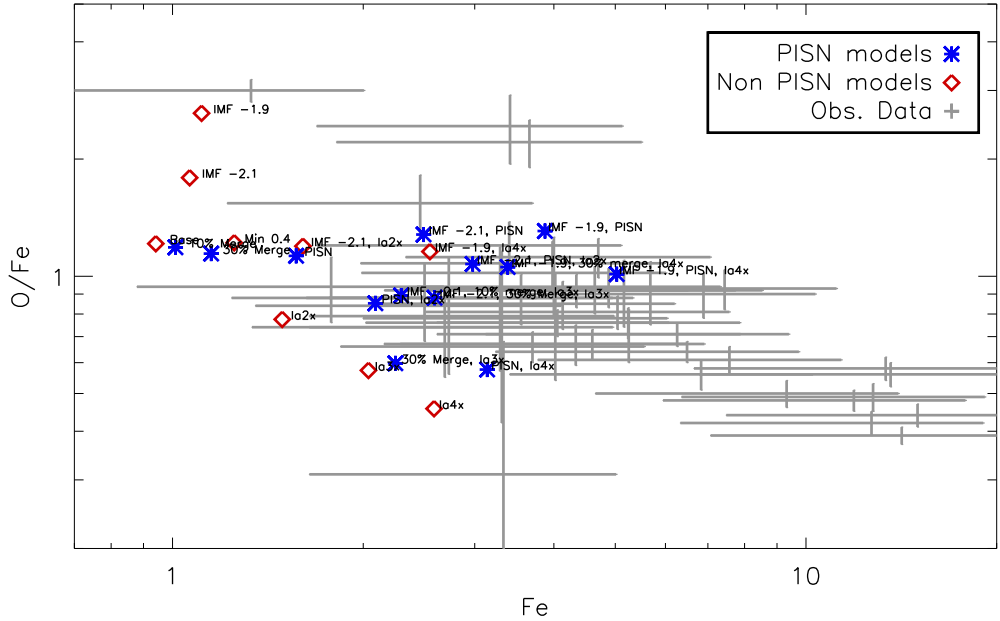}
\caption{
Increase in O/Fe ratio vs. total Fe abundance compared to solar values.
Data (grey points) are from \citet{lovisari11} for ICM observations of 5 galaxy clusters at different radii.
The Fe values and error bars assume stars account for 10\% of cluster baryons.
Models with no PISNe are shown as red diamonds.
Models with PISNe  are blue stars.
At least some of both types of models can produce reasonable values of O/Fe and total Fe, although only models with PISNe can produce very high Fe values (above 3).
}
\label{fig:Fe_O_vs_Fe}
\end{figure*}

In fig.~\ref{fig:Fe_O_vs_Fe} we plot the O/Fe ratio vs. the total amount of Fe produced.  
Observed and modeled values are relative to solar.
For the total amount of Fe produced for the observations (gray points) from \citet{lovisari11} we assume stars account for 10\% of the baryons in the cluster, and so multiply the observed ICM Fe abundance by a factor of 10, with error bars ranging from a factor of 5 to 15.
This range should accommodate variations of stellar to total baryon mass fractions, gas recycling, and metal contained in stars.
Several models with PISN (blue points) and without PISN (red points) can account for a depletion of O/Fe, particularly for models with an increased type Ia SN rate.
Models with only a flattened IMF instead produce in increase in O/Fe.
For models without any PISN, the highest amounts of Fe are produced with the highest type Ia SN rates.
However, even with 4 times the type Ia rate, only about 2.7 times as much iron is produced.
Models that include PISN have an easier time producing extra iron, reaching up to 5 times the base case for the most extreme model considered here.

\begin{figure*}
\includegraphics[scale=.9]{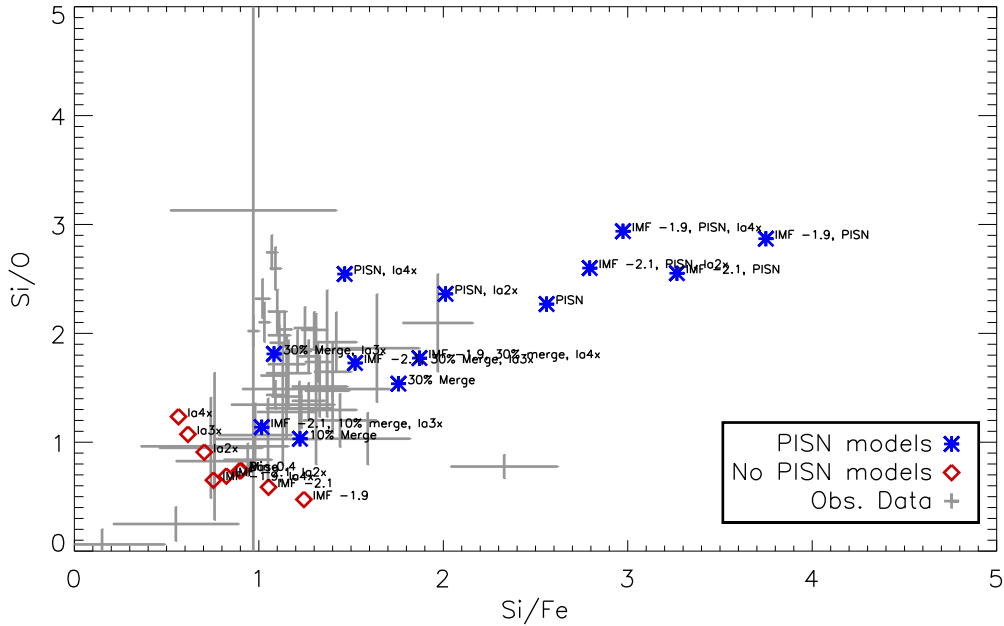}
\caption{
Increase in Si/O ratio vs. Si/Fe ratio compared to solar values.
Data (grey points) are from \citet{lovisari11} for ICM observations of 5 galaxy clusters at different radii.
Models with no PISNe are shown as red diamonds.
Models with PISNe  are blue stars.
Only models that include at least some PISNe can produce an enhancement of both Si/O and Si/Fe.
Si is over-produced relative to O and Fe for PISN models that simply extend the upper-mass cutoff of the IMF slope to include PISNe.
Better fits are obtained for models with 10\% or 30\% mergers for high-mass stars.
}
\label{fig:Si_Fe_vs_Si_O}
\end{figure*}

Comparing intermediate elements to abundances of oxygen and iron shows a clear division between models.
Figure~\ref{fig:Si_Fe_vs_Si_O} plots the ratio of Si/O vs. Si/Fe.
Observations, shown in gray, generally show an enhancement of both ratios.
Models without PISN (red points) cannot produce an enhancement in both Si/O and Si/Fe.
They can have enhanced Si/O and depleted Si/Fe (for increased type Ia SNe rates), depleted Si/O and enhanced Si/Fe (for flatted IMF slopes), or approximately solar values of both (for mixtures of flat IMF and increase type Ia rates or for removal of low-mass stars).
Models that include at least some PISNe (blue points) all show at least some increase in both Si/O and Si/Fe.
This is because PISNe produce such a large amount of intermediate elements.
It is not possible to get an enhancement of intermediate elements relative to both oxygen and iron without including pair-instability supernovae.

Models that extend the power-law slope of the IMF through the PISN region appear to overproduce Si and other intermediate elements.
Better fits to observations are obtained for models with 10\% or 30\% of stars above 65 M$_{\sun}$ merging, or, equivalently, about 10\% to 30\% the number of PISN predicted for a simple extension of the IMF to higher masses.

Note that our models reflect the total abundance of each element, so contributions from stars, gas in the ISM of cluster galaxies, the intra-cluster medium, and dust needs to be included when making detailed comparisons between observations and models.
The relative abundances in each of these could be different, and measuring the abundance of a particular elements in all of them can be challenging.

\section{Alternate Supernova Model}
\label{sec:kobayashi}

As an alternative to the core-collapse supernova models of \citet{heger10}, in this section we consider models from \citet{kobayashi2006}.
\citet{kobayashi2006} covers only a limited range of progenitor masses (7 masses between 13 M$_{\sun}$ and 40 M$_{\sun}$) and assumes all explosions have an energy of $10^{51}$~erg, whereas the data used from \citet{heger10} had 120 masses from 10 M$_{\sun}$ and 100 M$_{\sun}$ and 7 energies from $0.3$ and $3.0 \times 10^{51}$~erg. 
However, the \citet{kobayashi2006} models cover progenitor metallicity fractions of 0.0, 0.001, 0.004 and 0.02, whereas \citet{heger10} only model zero metallicity stars.

The supernova yields for \citet{kobayashi2006} and \citet{heger10} disagree significantly.
Even at similar energies, masses and at zero metallicity, the \citet{kobayashi2006} models have a much higher yields per supernova.
The yields decrease somewhat with increasing metallicity, but even at solar metallicity the \citet{kobayashi2006} produce much more metal.
This is not, however, necessarily unrealistic, if the metal content of the Milky Way, and other typical spiral galaxies, is not saturated but has been mixed with a significant amount of pristine gas.

\begin{figure*}
\includegraphics[scale=.9]{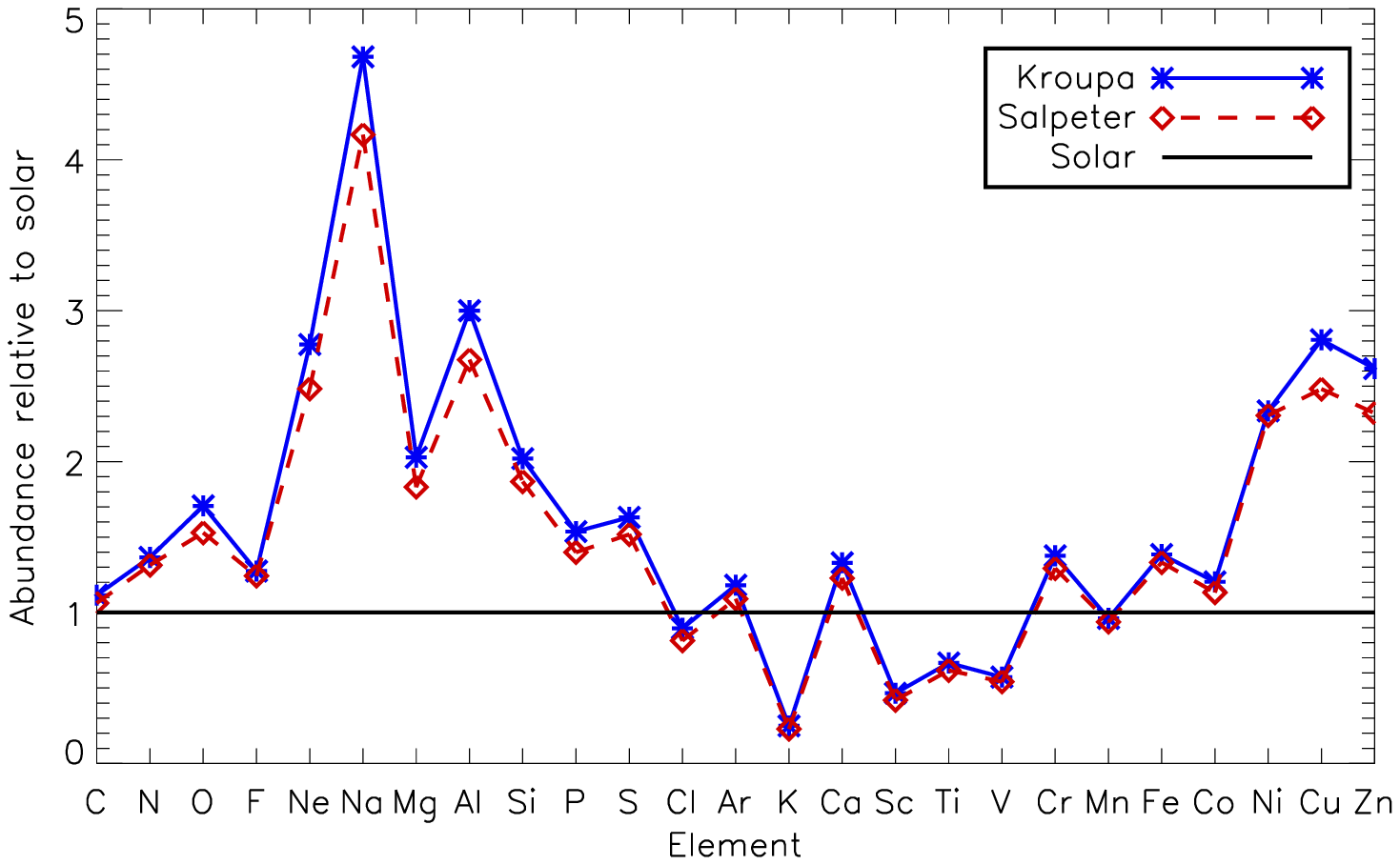}
\caption{
Same as fig.~\ref{fig:base_vs_solar} but using \citet{kobayashi2006} models for core-collapse supernovae.  
Mass abundance of each element from C to Zn relative to solar values for base models using a Kroupa IMF (solid blue line) and Salpeter IMF (dashed red line).
Solar abundances (solid black line) are all equal to 1 on this plot.
}
\label{fig:kobayashi_base_vs_solar}
\end{figure*}

To take a best case scenario using \citet{kobayashi2006} core-collapse supernovae models, we use only solar metallicity models.
Fig.~\ref{fig:kobayashi_base_vs_solar} plots the abundances of each elements compared to solar values.  
This is directly comparable to fig.~\ref{fig:base_vs_solar}.
The oxygen abundance is about 170\% solar, and iron is about 140\% solar.  
Very large enhancements are seen in lighter intermediate elements (Ne through Al).
Si and S are enhanced relative to solar, by a roughly similar amount as oxygen.

\begin{figure*}
\includegraphics[scale=.9]{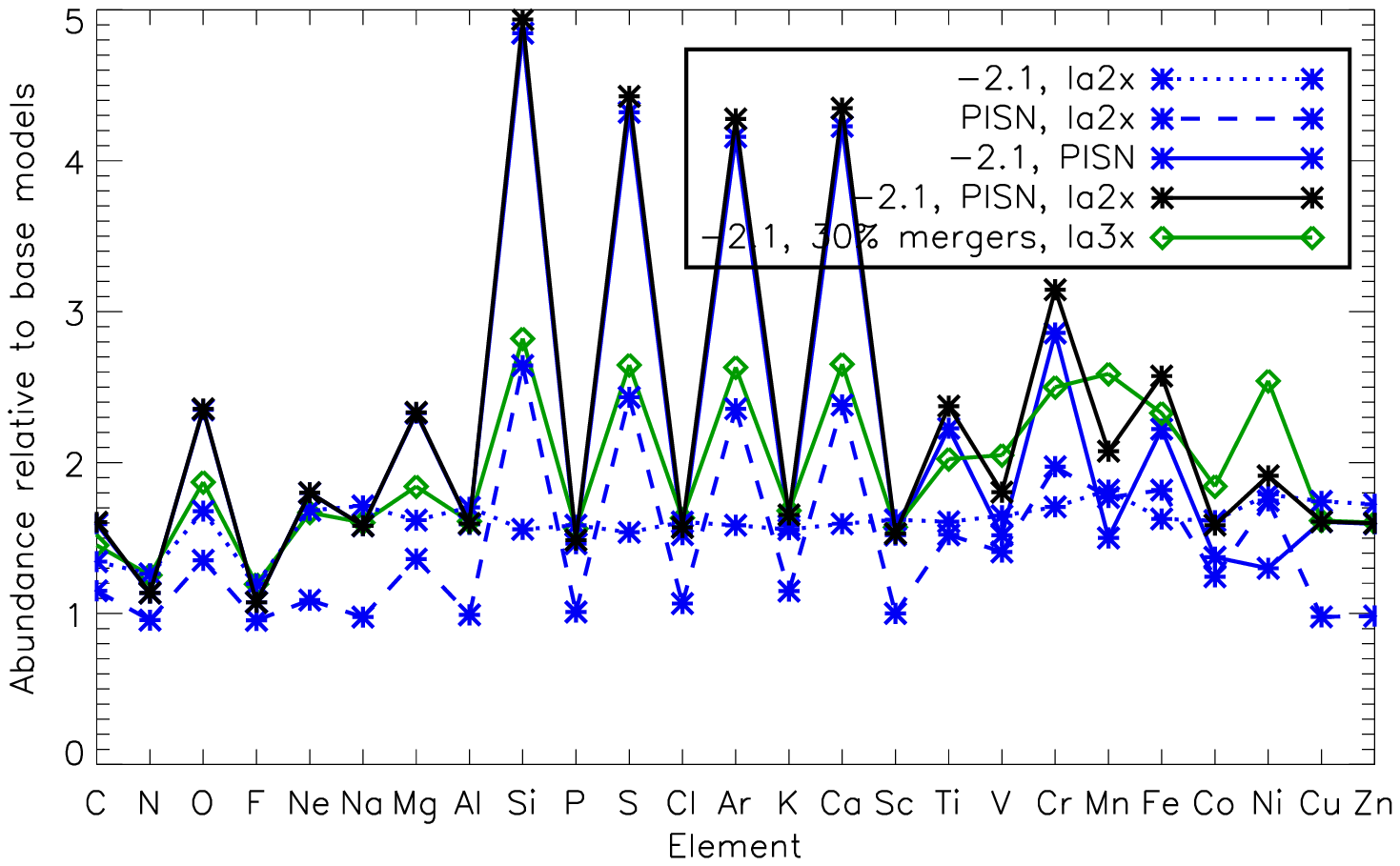}
\caption{
Same as fig.~\ref{fig:combined_vs_base} but using \citet{kobayashi2006} models for core-collapse supernovae.  
Mass abundance relative to base models for 5 combined models.  
Only models with a Kroupa-type IMF are shown.
All models that include PISNe show a clear signature in the enhancement of intermediate elements.
The largest increases in metal production are those with a flattened IMF and PISN (extended upper mass cutoff).
A compromise model (solid green line) with a flattened IMF, 30\% massive star merger rate, and 3x increased Type Ia SNe rate, produced similar increases in metal production to the PISN, 2x Type Ia rate model, but with more oxygen production.
}
\label{fig:kobayashi_combined_vs_base}
\end{figure*}

Because core-collapse supernova produce more intermediate elements, the increase of intermediate elements relative to oxygen and iron when PISNe are included is reduced.
For example, fig.~\ref{fig:kobayashi_combined_vs_base} compares the same scenarios as plotted in fig.~\ref{fig:combined_vs_base}, but using the \citet{kobayashi2006} core-collapse SNe models.  
Even in the most extreme cases, intermediate elements are only enhanced by an factor of 4 to 5 relative to the base model, rather than a factor of 6 to 10 when using the \citet{heger10} yields.

\begin{figure*}
\includegraphics[scale=.9]{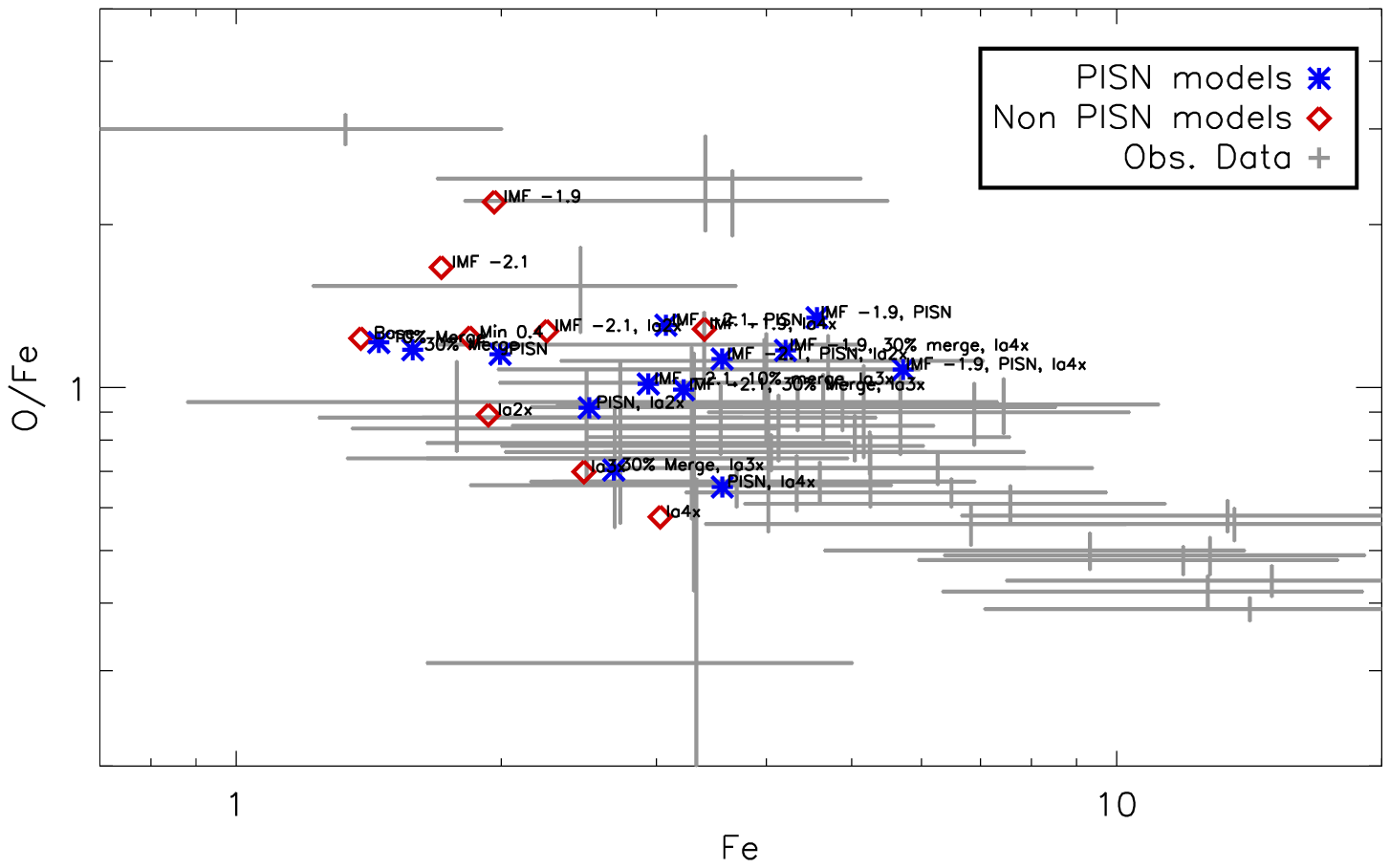}
\caption{
Same as fig.~\ref{fig:Fe_O_vs_Fe} but using \citet{kobayashi2006} models for core-collapse supernovae.  
Increase in O/Fe ratio vs. total Fe abundance compared to solar values.
Data (grey points) are from \citet{lovisari11} for ICM observations of 5 galaxy clusters at different radii.
The Fe values and error bars assume stars account for 10\% of cluster baryons.
Models with no PISNe are shown as red diamonds.
Models with PISNe  are blue stars.
At least some of both types of models can produce reasonable values of O/Fe and total Fe, although only models with PISNe can produce very high Fe values (above 3.5).
All models produce more Fe than the same models using \citet{heger10} core-collapse supernova models.
}
\label{fig:kobayashi_Fe_O_vs_Fe}
\end{figure*}

Due to the higher overall core-collapse SNe yields, the amount of metals produced in each model is higher as well.  
Fig.~\ref{fig:kobayashi_Fe_O_vs_Fe} plots the O/Fe ratio vs. the total amount of Fe produced, comparable to fig.~\ref{fig:Fe_O_vs_Fe}.
The O/Fe ratios are still reasonable, but the amount of iron produced is higher, making it easier to achieve the high metallicities seen in clusters.

\begin{figure*}
\includegraphics[scale=.9]{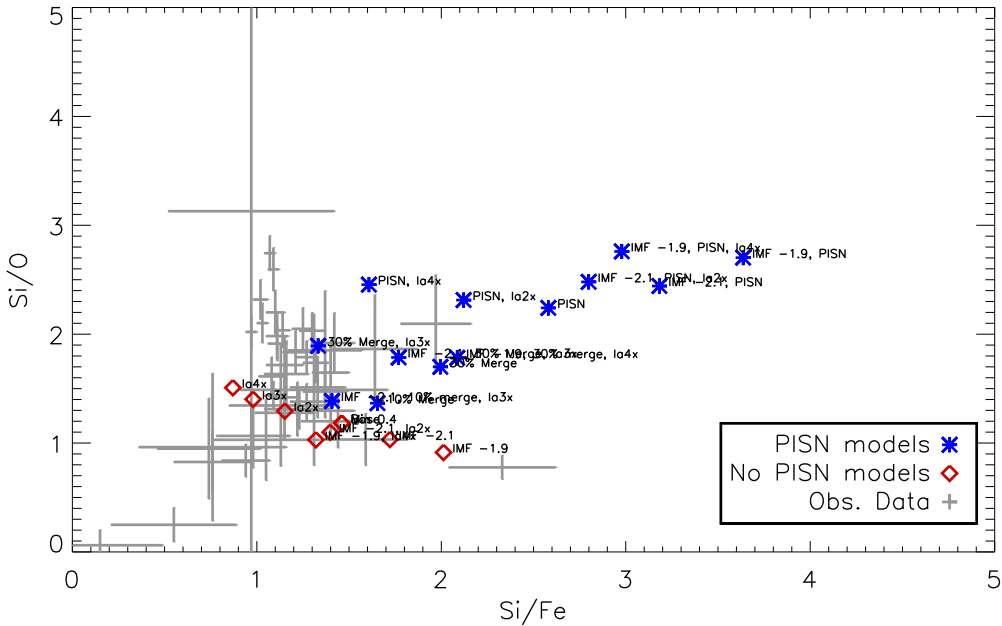}
\caption{
Same as fig.~\ref{fig:Si_Fe_vs_Si_O} but using \citet{kobayashi2006} models for core-collapse supernovae.  
Increase in Si/O ratio vs. Si/Fe ratio compared to solar values.
Data (grey points) are from \citet{lovisari11} for ICM observations of 5 galaxy clusters at different radii.
Models with no PISNe are shown as red diamonds.
Models with PISNe  are blue stars.
The Si/Fe ratio in the base model is significantly enhance relative to solar values.
However, only models that include at least some PISNe can produce an enhancement of both Si/O and Si/Fe relative to the base model.
Si is over-produced relative to O and Fe for PISN models that simply extend the upper-mass cutoff of the IMF slope to include PISNe.
Better fits are obtained for models with 10\% or 30\% mergers for high-mass stars.
}
\label{fig:kobayashi_Si_Fe_vs_Si_O}
\end{figure*}

In fig.~\ref{fig:kobayashi_Si_Fe_vs_Si_O} we plot the ratio of Si/O vs Si/Fe for the \citet{kobayashi2006} models, comparable to fig.~\ref{fig:Si_Fe_vs_Si_O}.
For the base model and others without PISNe (red diamonds), the ratio of Si/O and Si/Fe is enhanced, due to the increased production of Si in core-collapse SNe.
Note, however, that this is because the base model is out of line with solar abundances.
A higher rate of type Ia supernovae, particularly the Ia 2x model, produces close to solar abundance ratios.
Models with PISNe (blue stars) show a significant enhancement of Si relative to both O and Fe, and models with 10\% or 30\% merger rates still reproduce the cluster observations well.

The uncertainty in the yields from core-collapse supernovae can clearly have a significant effect on the overall metal production.
Higher yields, in particular, make it easier to produce the large amounts of metal seen in clusters.
Ideally, better supernova models, covering a wide range of progenitor mass, metallicity, and supernova energy would be used, but as of now \citet{heger10} and \citet{kobayashi2006} are the best data sets available.  
Although they significantly disagree with each other about the metal yield per supernova, even for nearly identical conditions, the choice of model does not qualitatively change our results.
Pair-instability supernova still significantly enhance total metal production, and are necessary to simultaneously increase the ratio of intermediate elements to oxygen and iron.

\section{Conclusions}
\label{sec:conclusions}

Based on our models, we are able to draw the following conclusions:

1. Including contributions from pair-instability supernova significantly increases total metal production.
This is particularly true when combined with an increased type Ia SN rate and a slightly flattened IMF slope.
The PISNe can be produced either by increasing the upper mass cutoff of the IMF or by mergers between stars.
A spatially or temporally distinct population of high-mass stars, e.g. a very top-heavy IMF for intra-cluster stars or a large amount of Population III star formation, is not necessary to produce the high amounts of metal found in galaxy clusters.

2. PISN are necessary to simultaneously produce enhancements of intermediate-mass elements relative to oxygen and iron.  
Each means of increasing metal production in clusters create a distinct abundance signature.
An enhancement of intermediate elements relative to oxygen can be produced by excess type Ia SNe, but with a depletion relative to iron.  
The opposite will occur if there is an excess of core-collapse SNe (i.e. for a flattened IMF).
Models with an excess of both core-collapse and Type Ia SNe, or with low-mass stars removed, will produce approximately solar abundance ratios.
Only by including PISN can the amount of intermediate elements be increased relative to both oxygen and iron.

3. Models that simply extend the high-mass slope of the IMF to a cutoff mass above the PISN range over produce intermediate elements relative to oxygen and iron.
Models with PISNe from mergers of 10\% to 30\% of stars above 65 M$_{\sun}$ appear to produce better fits.
This could indicate that there are fewer stars then expected for a continuous IMF in the right mass range (140 to 260 M$_{\sun}$), that the upper mass cutoff of the IMF is still below 260 M$_{\sun}$ in galaxy clusters, or that an initial mass higher than 140 M$_{\sun}$ is required to produce a PISN.
It could also be that not all galaxies in clusters produce PISN.
For example, it may be that only massive galaxies or only galaxies near the center of the cluster are able to produce stars large enough for a PISN to occur.

4. We expect 1 PISN to occur for every 60 to 600 core-collapse supernovae in galaxy clusters. 
For a cluster with a star formation rate of 100 M$_{\sun}$/yr, this would translate to 1 PISN every 20 to 600 years, depending on the IMF.
This assumes the number of PISN per stellar mass formed is constant over the life of a cluster.
It is possible PISN could only occur at and before the time of peak star formation in the cluster, when average metallicity would have been lower.
This would, however, require more PISNe per stellar mass formed at these times.

The largest uncertainties in our models come from supernova simulations results used as input.  
None of the simulations of type Ia, core-collapse, or pair-instability supernovae are completely realistic, e.g., none are 3D simulations, none naturally produce a SN, etc.
Better models might change the amounts of metals produced by different supernovae, or change the relative amounts of specific elements, but so long at the general trends are the same this should not qualitatively change our results.
We also do not consider the metallicity history of the cluster galaxies in our models, aside from using a fixed range of metallicities for AGB stars.

It is also unknown under what conditions PISNe will be created.  
The simulation results we use \citep{heger02} are for non-rotating stars with an initial metallicity of zero.
Stars with non-zero metallicity may loose too much mass for a PISN to occur, or the initial mass lower-limit for PISN may be higher.
Alternatively, more recent work \citep{chatzopoulos12} has shown that PISN can occur a much lower mass ($\approx 65$~M$_{\sun}$ rather than $140$~M$_{\sun}$) for rapidly-rotating stars.
Real PISNe may occur at different masses and produce very different amounts of each element that the values we assume here.
In particular, the abundance of intermediate-mass elements is very sensitive to the number of PISNe in our models, and is therefore very dependent on the PISNe models we use.
However, so long as the general result that PISNe produce enhanced amounts of intermediate elements relative to oxygen and iron remains true, the conclusion that PISNe are necessary in galaxy clusters will remain true.

\section*{Acknowledgments}

BJM is supported by an NSF Astronomy and Astrophysics Postdoctoral Fellowship under award AST1102796.
The authors thank Michael Loewenstein for his many useful comments.

\bibliographystyle{mn2e}
\bibliography{references}

\begin{thebibliography}{}

\bibitem[\protect\citeauthoryear{{Aguirre}, {Schaye}, {Kim}, {Theuns}, {Rauch}
  \& {Sargent}}{{Aguirre} et~al.}{2004}]{aguirre04}
{Aguirre} A.,  {Schaye} J.,  {Kim} T.-S.,  {Theuns} T.,  {Rauch} M.,
  {Sargent} W.~L.~W.,  2004, \apj, 602, 38

\bibitem[\protect\citeauthoryear{{Anders} \& {Grevesse}}{{Anders} \&
  {Grevesse}}{1989}]{anders89}
{Anders} E.,  {Grevesse} N.,  1989, \gca, 53, 197

\bibitem[\protect\citeauthoryear{{Andreon}}{{Andreon}}{2010}]{andreon10}
{Andreon} S.,  2010, \mnras, 407, 263

\bibitem[\protect\citeauthoryear{{Arnaud}, {Pratt}, {Piffaretti},
  {B{\"o}hringer}, {Croston} \& {Pointecouteau}}{{Arnaud}
  et~al.}{2010}]{arnaud10}
{Arnaud} M.,  {Pratt} G.~W.,  {Piffaretti} R.,  {B{\"o}hringer} H.,  {Croston}
  J.~H.,    {Pointecouteau} E.,  2010, \aap, 517, A92

\bibitem[\protect\citeauthoryear{{Auger}, {Treu}, {Bolton}, {Gavazzi},
  {Koopmans}, {Marshall}, {Moustakas} \& {Burles}}{{Auger}
  et~al.}{2010}]{auger10}
{Auger} M.~W.,  {Treu} T.,  {Bolton} A.~S.,  {Gavazzi} R.,  {Koopmans}
  L.~V.~E.,  {Marshall} P.~J.,  {Moustakas} L.~A.,    {Burles} S.,  2010, \apj,
  724, 511

\bibitem[\protect\citeauthoryear{{Banerjee}, {Kroupa} \& {Oh}}{{Banerjee}
  et~al.}{2012}]{banerjee12}
{Banerjee} S.,  {Kroupa} P.,    {Oh} S.,  2012, \mnras, 426, 1416

\bibitem[\protect\citeauthoryear{{Bregman}, {Anderson} \& {Dai}}{{Bregman}
  et~al.}{2010}]{bregman10}
{Bregman} J.~N.,  {Anderson} M.~E.,    {Dai} X.,  2010, \apjl, 716, L63

\bibitem[\protect\citeauthoryear{{Canning}, {Fabian}, {Johnstone}, {Sanders},
  {Conselice}, {Crawford}, {Gallagher} \& {Zweibel}}{{Canning}
  et~al.}{2010}]{canning10}
{Canning} R.~E.~A.,  {Fabian} A.~C.,  {Johnstone} R.~M.,  {Sanders} J.~S.,
  {Conselice} C.~J.,  {Crawford} C.~S.,  {Gallagher} J.~S.,    {Zweibel} E.,
  2010, \mnras, 405, 115

\bibitem[\protect\citeauthoryear{{Chatzopoulos} \& {Wheeler}}{{Chatzopoulos} \&
  {Wheeler}}{2012}]{chatzopoulos12}
{Chatzopoulos} E.,  {Wheeler} J.~C.,  2012, \apj, 748, 42

\bibitem[\protect\citeauthoryear{{Crowther}, {Schnurr}, {Hirschi}, {Yusof},
  {Parker}, {Goodwin} \& {Kassim}}{{Crowther} et~al.}{2010}]{crowther10}
{Crowther} P.~A.,  {Schnurr} O.,  {Hirschi} R.,  {Yusof} N.,  {Parker} R.~J.,
  {Goodwin} S.~P.,    {Kassim} H.~A.,  2010, \mnras, 408, 731

\bibitem[\protect\citeauthoryear{{Edge} \& {Stewart}}{{Edge} \&
  {Stewart}}{1991}]{edge91}
{Edge} A.~C.,  {Stewart} G.~C.,  1991, \mnras, 252, 428

\bibitem[\protect\citeauthoryear{{Ellison}, {Simard}, {Cowan}, {Baldry},
  {Patton} \& {McConnachie}}{{Ellison} et~al.}{2009}]{ellison09}
{Ellison} S.~L.,  {Simard} L.,  {Cowan} N.~B.,  {Baldry} I.~K.,  {Patton}
  D.~R.,    {McConnachie} A.~W.,  2009, \mnras, 396, 1257

\bibitem[\protect\citeauthoryear{{Espinoza}, {Selman} \& {Melnick}}{{Espinoza}
  et~al.}{2009}]{espinoza09}
{Espinoza} P.,  {Selman} F.~J.,    {Melnick} J.,  2009, \aap, 501, 563

\bibitem[\protect\citeauthoryear{{Fabjan}, {Tornatore}, {Borgani}, {Saro} \&
  {Dolag}}{{Fabjan} et~al.}{2008}]{fabjan08}
{Fabjan} D.,  {Tornatore} L.,  {Borgani} S.,  {Saro} A.,    {Dolag} K.,  2008,
  \mnras, 386, 1265

\bibitem[\protect\citeauthoryear{{Gal-Yam} et~al.,}{{Gal-Yam}
  et~al.}{2009}]{galyam09}
{Gal-Yam} A.  et~al., 2009, \nat, 462, 624

\bibitem[\protect\citeauthoryear{{Gonzalez}, {Zaritsky} \&
  {Zabludoff}}{{Gonzalez} et~al.}{2007}]{gonzalez07}
{Gonzalez} A.~H.,  {Zaritsky} D.,    {Zabludoff} A.~I.,  2007, \apj, 666, 147

\bibitem[\protect\citeauthoryear{{Harayama}, {Eisenhauer} \&
  {Martins}}{{Harayama} et~al.}{2008}]{harayama08}
{Harayama} Y.,  {Eisenhauer} F.,    {Martins} F.,  2008, \apj, 675, 1319

\bibitem[\protect\citeauthoryear{{Harris} \& {Racine}}{{Harris} \&
  {Racine}}{1979}]{harris91}
{Harris} W.~E.,  {Racine} R.,  1979, \araa, 17, 241

\bibitem[\protect\citeauthoryear{{Heger} \& {Woosley}}{{Heger} \&
  {Woosley}}{2002}]{heger02}
{Heger} A.,  {Woosley} S.~E.,  2002, \apj, 567, 532

\bibitem[\protect\citeauthoryear{{Heger} \& {Woosley}}{{Heger} \&
  {Woosley}}{2010}]{heger10}
{Heger} A.,  {Woosley} S.~E.,  2010, \apj, 724, 341

\bibitem[\protect\citeauthoryear{{Iwamoto}, {Brachwitz}, {Nomoto}, {Kishimoto},
  {Umeda}, {Hix} \& {Thielemann}}{{Iwamoto} et~al.}{1999}]{iwamoto99}
{Iwamoto} K.,  {Brachwitz} F.,  {Nomoto} K.,  {Kishimoto} N.,  {Umeda} H.,
  {Hix} W.~R.,    {Thielemann} F.-K.,  1999, \apjs, 125, 439

\bibitem[\protect\citeauthoryear{{Karakas}}{{Karakas}}{2010}]{karakas10}
{Karakas} A.~I.,  2010, \mnras, 403, 1413

\bibitem[\protect\citeauthoryear{{Kobayashi}, {Umeda}, {Nomoto}, {Tominaga} \&
  {Ohkubo}}{{Kobayashi} et~al.}{2006}]{kobayashi2006}
{Kobayashi} C.,  {Umeda} H.,  {Nomoto} K.,  {Tominaga} N.,    {Ohkubo} T.,
  2006, \apj, 653, 1145

\bibitem[\protect\citeauthoryear{{Krick} \& {Bernstein}}{{Krick} \&
  {Bernstein}}{2007}]{krick07}
{Krick} J.~E.,  {Bernstein} R.~A.,  2007, \aj, 134, 466

\bibitem[\protect\citeauthoryear{{Loewenstein}}{{Loewenstein}}{2001}]{loewenstein01}
{Loewenstein} M.,  2001, \apj, 557, 573

\bibitem[\protect\citeauthoryear{{Loewenstein}}{{Loewenstein}}{2006}]{loewenstein06}
{Loewenstein} M.,  2006, \apj, 648, 230

\bibitem[\protect\citeauthoryear{{Loewenstein}}{{Loewenstein}}{2013}]{loewenstein13}
{Loewenstein} M.,  2013, \apj, 773, 52

\bibitem[\protect\citeauthoryear{{Lovisari}, {Schindler} \&
  {Kapferer}}{{Lovisari} et~al.}{2011}]{lovisari11}
{Lovisari} L.,  {Schindler} S.,    {Kapferer} W.,  2011, \aap, 528, A60

\bibitem[\protect\citeauthoryear{{Mannucci}, {Maoz}, {Sharon}, {Botticella},
  {Della Valle}, {Gal-Yam} \& {Panagia}}{{Mannucci} et~al.}{2008}]{mannucci08}
{Mannucci} F.,  {Maoz} D.,  {Sharon} K.,  {Botticella} M.~T.,  {Della Valle}
  M.,  {Gal-Yam} A.,    {Panagia} N.,  2008, \mnras, 383, 1121

\bibitem[\protect\citeauthoryear{{Maoz} \& {Mannucci}}{{Maoz} \&
  {Mannucci}}{2012}]{maoz12}
{Maoz} D.,  {Mannucci} F.,  2012, PASA, 29, 447

\bibitem[\protect\citeauthoryear{{Massey} \& {Hunter}}{{Massey} \&
  {Hunter}}{1998}]{massey98}
{Massey} P.,  {Hunter} D.~A.,  1998, \apj, 493, 180

\bibitem[\protect\citeauthoryear{{McDonald} et~al.,}{{McDonald}
  et~al.}{2012}]{mcdonald12}
{McDonald} M.  et~al., 2012, \nat, 488, 349

\bibitem[\protect\citeauthoryear{{McNamara} \& {O'Connell}}{{McNamara} \&
  {O'Connell}}{1989}]{mcnamara89}
{McNamara} B.~R.,  {O'Connell} R.~W.,  1989, \aj, 98, 2018

\bibitem[\protect\citeauthoryear{{Mushotzky} \& {Loewenstein}}{{Mushotzky} \&
  {Loewenstein}}{1997}]{mushotzky97}
{Mushotzky} R.~F.,  {Loewenstein} M.,  1997, \apjl, 481, L63

\bibitem[\protect\citeauthoryear{{O'Dea} et~al.,}{{O'Dea}
  et~al.}{2010}]{odea10}
{O'Dea} K.~P.  et~al., 2010, \apj, 719, 1619

\bibitem[\protect\citeauthoryear{{Pagel}}{{Pagel}}{1997}]{pagel97}
{Pagel} B.~E.~J.,  1997, {Nucleosynthesis and Chemical Evolution of Galaxies}.
Cambridge University Press

\bibitem[\protect\citeauthoryear{{Portinari}, {Moretti}, {Chiosi} \&
  {Sommer-Larsen}}{{Portinari} et~al.}{2004}]{portinari04}
{Portinari} L.,  {Moretti} A.,  {Chiosi} C.,    {Sommer-Larsen} J.,  2004,
  \apj, 604, 579

\bibitem[\protect\citeauthoryear{{Sana} et~al.,}{{Sana} et~al.}{2013}]{sana13}
{Sana} H.  et~al., 2013, \aap, 550, A107

\bibitem[\protect\citeauthoryear{{Sand} et~al.,}{{Sand} et~al.}{2012}]{sand12}
{Sand} D.~J.  et~al., 2012, \apj, 746, 163

\bibitem[\protect\citeauthoryear{{Sanders} \& {Fabian}}{{Sanders} \&
  {Fabian}}{2006}]{sanders06}
{Sanders} J.~S.,  {Fabian} A.~C.,  2006, \mnras, 371, 1483

\bibitem[\protect\citeauthoryear{{Sharon}, {Gal-Yam}, {Maoz}, {Filippenko} \&
  {Guhathakurta}}{{Sharon} et~al.}{2007}]{sharon07}
{Sharon} K.,  {Gal-Yam} A.,  {Maoz} D.,  {Filippenko} A.~V.,    {Guhathakurta}
  P.,  2007, \apj, 660, 1165

\bibitem[\protect\citeauthoryear{{Tamura}, {Sharples}, {Arimoto}, {Onodera},
  {Ohta} \& {Yamada}}{{Tamura} et~al.}{2006}]{tamura06}
{Tamura} N.,  {Sharples} R.~M.,  {Arimoto} N.,  {Onodera} M.,  {Ohta} K.,
  {Yamada} Y.,  2006, \mnras, 373, 588

\bibitem[\protect\citeauthoryear{{Tang}, {Gu} \& {Huang}}{{Tang}
  et~al.}{2009}]{tang09}
{Tang} B.-T.,  {Gu} Q.-S.,    {Huang} S.,  2009, Research in Astronomy and
  Astrophysics, 9, 1215

\bibitem[\protect\citeauthoryear{{Tinsley}}{{Tinsley}}{1980}]{tinsley80}
{Tinsley} B.~M.,  1980, \fcp, 5, 287

\bibitem[\protect\citeauthoryear{{Trager}, {Faber}, {Worthey} \&
  {Gonz{\'a}lez}}{{Trager} et~al.}{2000a}]{trager00b}
{Trager} S.~C.,  {Faber} S.~M.,  {Worthey} G.,    {Gonz{\'a}lez} J.~J.,  2000a,
  \aj, 120, 165

\bibitem[\protect\citeauthoryear{{Trager}, {Faber}, {Worthey} \&
  {Gonz{\'a}lez}}{{Trager} et~al.}{2000b}]{trager00}
{Trager} S.~C.,  {Faber} S.~M.,  {Worthey} G.,    {Gonz{\'a}lez} J.~J.,  2000b,
  \aj, 119, 1645

\bibitem[\protect\citeauthoryear{{Treu}, {Auger}, {Koopmans}, {Gavazzi},
  {Marshall} \& {Bolton}}{{Treu} et~al.}{2010}]{treu10}
{Treu} T.,  {Auger} M.~W.,  {Koopmans} L.~V.~E.,  {Gavazzi} R.,  {Marshall}
  P.~J.,    {Bolton} A.~S.,  2010, \apj, 709, 1195

\bibitem[\protect\citeauthoryear{{van Dokkum} \& {Conroy}}{{van Dokkum} \&
  {Conroy}}{2010}]{vandokkum10}
{van Dokkum} P.~G.,  {Conroy} C.,  2010, \nat, 468, 940

\bibitem[\protect\citeauthoryear{{Weidner}, {Ferreras}, {Vazdekis} \& {La
  Barbera}}{{Weidner} et~al.}{2013}]{weidner13a}
{Weidner} C.,  {Ferreras} I.,  {Vazdekis} A.,    {La Barbera} F.,  2013,
  \mnras, 435, 2274

\bibitem[\protect\citeauthoryear{{Weidner}, {Kroupa} \&
  {Pflamm-Altenburg}}{{Weidner} et~al.}{2013}]{weidner13}
{Weidner} C.,  {Kroupa} P.,    {Pflamm-Altenburg} J.,  2013, \mnras, 434, 84

\bibitem[\protect\citeauthoryear{{Yamashita}}{{Yamashita}}{1992}]{yamashita92}
{Yamashita} K.,  1992, in {Tanaka} Y.,  {Koyama} K.,  eds, Frontiers Science
  Series. p.~475

\end{thebibliography}


\end{document}